\documentclass[aps,twocolumn,groupedaddress,showpacs]{revtex4}
\usepackage{graphicx}
\begin{document}
\bibliographystyle{apsrev}

\draft
\preprint{}
\title{Magnetic Order in YBa$_2$Cu$_3$O$_{6+x}$ Superconductors}
\author{H. A. Mook,$^1$ Pengcheng Dai,$^{2,1}$
S. M. Hayden,$^3$ A. Hiess,$^{4}$ J. W. Lynn,$^5$ S.-H. Lee,$^{5}$ and F. Do$\rm\breve{g}$an$^{6}$}
\address{$^1$Solid State Division, Oak Ridge National Laboratory, Oak Ridge, Tennessee 37831-6393}
\address{$^2$ Department of Physics and Astronomy, The University of Tennessee, Knoxville, Tennessee 37996}
\address{$^3$ H. H. Wills Physics Laboratory, University of Bristol, Bristol BS8 1TL, UK}
\address{$^4$ Institut Laue-Langevin, BP 156, 38042 Grenoble, France}
\address{$^5$ NIST Center for Neutron Research, National Institute of Standards and Technology, 
Gaithersburg, Maryland 20899}
\address{$^6$ Department of Materials Science and Engineering, University of Washington, Seattle, WA 98195}

\date{\today}
\begin{abstract}
Polarized and unpolarized neutron diffraction has been used to search for magnetic order in YBa$_2$Cu$_3$O$_{6+x}$ superconductors. Most of the measurements were made on a high quality crystal of YBa$_2$Cu$_3$O$_{6.6}$. 
It is shown that this crystal has highly ordered ortho-II chain order, and a sharp superconducting transition. Inelastic scattering measurements display a very clean spin-gap and pseudogap 
with any intensity at 10 meV being 50 times smaller than the resonance intensity. The crystal shows a complicated magnetic order that appears to have three components. A magnetic phase is found at high temperatures that seems to stem from an impurity with a moment that is in the $a$-$b$ plane, but disordered on the crystal lattice. A second ordering occurs near the pseudogap temperature that has a shorter correlation length 
than the high temperature phase and a moment direction that is at least partly along the $c$-axis of the crystal. Its moment direction, temperature dependence, and Bragg intensities suggest that it may stem from orbital ordering of the $d$-density wave (DDW) type.  An additional intensity increase occurs below the superconducting transition. The magnetic intensity in these phases does not change noticeably in a 7 Tesla magnetic field aligned 
approximately along the $c$-axis. Searches for magnetic order in YBa$_2$Cu$_3$O$_{7}$ show no signal while a small magnetic intensity is found in YBa$_2$Cu$_3$O$_{6.45}$ that is consistent with $c$-axis directed magnetic order. The results are contrasted with other recent neutron measurements. 
\end{abstract}

\maketitle

\narrowtext

\section{Introduction}
It is well understood that superconductivity produces a gap in the quasiparticle spectra, and in conventional materials this gap disappears as the temperature is increased to $T_c$ where the superconducting electron pairs are no longer bound together.  The problem is that for the underdoped cuprate materials this gap appears to get bigger as $T_c$ gets smaller. This pseudogap is one of the most puzzling attributes of the cuprate superconductors. A news article on the pseudogap can be found in Ref. \cite{buchanan} while a comprehensive review of the various experimental techniques used to determine the pseudogap is given by Timusk and Statt \cite{timusk}. One way to understand the pseudogap is to postulate that phase-incoherent pairs are established as the material is cooled through the pseudogap temperature $T^\ast$ and superconductivity takes place at a lower temperature $T_c$ 
when phase coherence is established \cite{emery1,uemura}. An experiment on the sample used in the present study suggested that pre-formed pairs were present in the normal state that leads to superconductivity at a reduced temperature \cite{dain}. It appeared that preformed pairs might not be present at a temperature 
as high as $T^\ast$, but this could not be conclusively established.

A completely different approach to the problem was taken by Varma \cite{varma}, who postulated a state with broken time and rotational symmetry that displayed a pattern of circulating currents (CC phase) in the $a$-$b$ plane.  
This state appeared at $T^\ast$ and for the near optimal doping case ended at a quantum critical point.  The currents of the CC phase would produce a signal visible by a sufficiently sensitive neutron scattering experiment at certain $(h, k, l)$ positions in the reciprocal lattice.  Observing these peaks necessitates a difficult polarized beam experiment that was undertaken \cite{lee}, but no effect could be found.  However, if the scattering was largely two-dimensional as in the present case, it is unlikely the signal would be observed.  Circulating currents around the Cu-O bonds that resulted from the staggered flux phase of the $t$-$J$ model were introduced 
by Hsu {\it et al.} \cite{hsu} and a neutron scattering experiment was considered to observe such a phase.  Chakravarty {\it et al.} \cite{chakravarty1} recently proposed a picture in some respects similar to the Varma proposal, but with physical currents that circulate around the Cu-O bonds.  This state competes with superconductivity and provides an explanation of the relationship between the pseudogap and superconductivity.  In this case a new state, termed the $d$-density wave (DDW), is formed below $T^\ast$.  This state breaks translational symmetry as well as time reversal and rotational symmetry.  The moments from the bond currents result in peaks at the $(h/2, k/2, l)$ superlattice positions of the reciprocal lattice that can be detected by a neutron scattering experiment of sufficient sensitivity.  A phase with the same current paths was considered by 
Ivanov {\it et al.} \cite{ivanov}, but in this case the moments fluctuate and can only be 
observed in vortex cores by neutrons.

This paper describes experiments designed to search for the DDW state. Most of the measurements were made on a high quality crystal of YBa$_2$Cu$_3$O$_{6.6}$  that has a hole doping of about 
0.1 electron per planar Cu and a $T_c$ of 63 K.  A preliminary experiment has been done on the 
YBa$_2$Cu$_3$O$_{6.6}$ crystal and a small signal was observed that became established below about 200 K 
and increased below $T_c$ \cite{mook1}. However, the experiment was done without polarized neutrons so that only limited information could be obtained. A polarized neutron study has now been completed that shows a similar temperature dependence for the magnetic intensity as is shown in Fig.\ 1. 
A high temperature phase (region $C$), is found above about 200 K, while the same increase as was observed 
before is found below 200 K (region $B$), with a further increase below $T_c$ (region $A$).  
The region of interest related to the pseudogap is region $B$ and the paper will discuss this 
scattering in detail and show how it differs from the scattering in region $C$. The scattering in 
region $A$ has not been investigated in detail, but doesn't seem to differ in any major way from the scattering in region $B$. An experiment has been done by Sidis {\it et al.} \cite{sidis} on a 
crystal of  YBa$_2$Cu$_3$O$_{6.5}$ ($T_c = 55$ K) that 
shows a rather large magnetic signal with a high temperature phase and an increase in intensity 
below $T_c$.  This experiment at first sight seems in conflict with the present work, but 
on further examination appears to be consistent with it. A very recent experiment by 
Stock {\it et al.} \cite{stock} performed on a high quality crystal of YBa$_2$Cu$_3$O$_{6.5}$ 
gave no indication of a magnetic signal. A sensitivity of 0.003 $\mu_B$ is claimed
if the signal was as sharp as the resolution width, 
but as will be  shown signals of that size are very difficult to observe without polarization 
analysis. This crystal has a large magnetic background that increases considerably upon cooling 
that is not observed in any of our crystals. There is also no observation of a pseudogap  for energies as low as 6.2 meV. It is thus questionable if the crystal might be expected to display the DDW state. 

\begin{figure}
\includegraphics[keepaspectratio=true, totalheight = 3.0 in, width = 2.0 in]{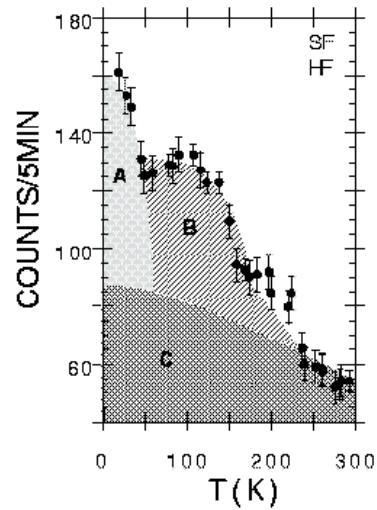}
\caption{Temperature dependence of the magnetic scattering from YBa$_2$Cu$_3$O$_{6.6}$. 
The curve  appears to have three main regions as shown by the differently shaded patterns.
}
\end{figure}

\section{Polarization analysis }
Polarized neutrons have played a major role in understanding the nature of the observed magnetic signal, 
so we first discuss how polarization analysis is used to characterize the magnetic signal. The seminal paper on polarization analysis was published by Moon {\it et al.} in 1969 \cite{moon} . We will only discuss the issues of interest for the present experiment, but background information is available in this reference. Our polarized beam measurements were made on the IN20 spectrometer at the Institut Laue-Langevin neutron source in Grenoble, France. The spectrometer arrangement was the standard one used for polarization analysis and employed Heusler alloy crystals for the monochromator and analyzer.  The collimation was 40$^\prime$-40$^\prime$-40$^\prime$-60$^\prime$ 
from in front of the monochromator until after the analyzer.  The neutron energy was 13.78 meV providing an energy resolution of about 1 meV, and three pyrolytic graphite filters were used to avoid higher order contamination. 
The scattering plane used in the experiment is shown in Fig.\ 2. The $(h, h, l)$ zone was selected so that the reflections of interest for the DDW state could be examined. The crystals used in the experiments are twinned so that 
$a^\ast$ could not be differentiated from $b^\ast$. We thus consider the crystals to be tetragonal so that the 
$[1, 1, 0]$ direction is in the basal plane and the $[1, -1, 0]$ direction perpendicular to the 
page is also in the basal plane. We describe our reciprocal lattice positions by 
$|{\bf a^\ast}|\approx |{\bf b^\ast}| =\pi/(a+b)$ and $|{\bf c^\ast}|=2\pi/c$, and the scans to be shown are made in reciprocal lattice units (r.l.u.).   Many of our results were obtained using the $(1/2, 1/2, 1)$ reflection as shown in Fig.\ 2. The momentum transfer ${\bf Q}$ for this reflection is determined by the difference in the incoming  and outgoing neutron wavevectors, ${\bf k-k^\prime}$. Guide fields were used to provide a neutron polarization direction either vertical to the scattering plane (vertical field or VF) or along the ${\bf Q}$ (horizontal field or HF), and a standard spin-flipping coil was used in the scattered beam.  A flipping ratio of 19 was measured in both field configurations. All the data shown in the paper have been corrected for the small amount of incomplete polarization.  

\begin{figure}
\includegraphics[keepaspectratio=true, totalheight = 2.0 in, width = 2.0 in]{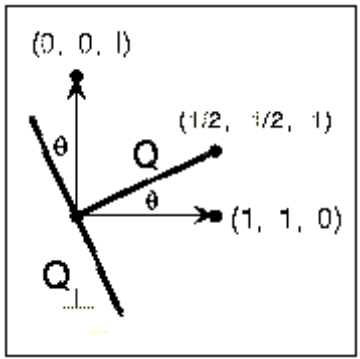}
\caption{
Scattering diagram for the polarized neutron experiment. The 
$(1/2, 1/2, 1)$ reflection occurs at an angle $\theta$ from the $[1, 1, 0]$ 
direction in the crystal. ${\bf Q}$ is the scattering vector for the reflection while the 
vector ${\bf Q}_\perp$ is in the scattering plane and is perpendicular to ${\bf Q}$.  
${\bf c}^\ast$  is along the $[0, 0, 1]$ direction which corresponds to the $c$-axis of the crystal.
}
\end{figure}

The magnetic neutron scattering intensity always originates from the moment projected on the plane 
perpendicular to ${\bf Q}$. The polarization analysis technique can further 
determine how much of the moment lies along the direction of ${\bf Q}_\perp$. The neutron polarization is directed either along or opposite to the guide field direction. We denote a scattering event in which the spin makes a 180 degree rotation as spin flip (SF), while scattering with no spin direction change is non-spin flip (NSF).  For the HF case all magnetic scattering is SF and stems from all the moment in the crystal projected on the  plane perpendicular to ${\bf Q}$. The NSF scattering arises from nonmagnetic processes. For the VF case the magnetic scattering is divided up into SF and NSF parts. The SF part of the magnetic scattering lies in the scattering along ${\bf Q}_\perp$. 
The NSF magnetic scattering stems from the moments pointing in the direction perpendicular to the scattering plane. If we adopt the notation of Sidis {\it et al.} \cite{sidis} the HF SF scattering for the $(1/2, 1/2, 1)$ 
reflection as shown on Fig.\ 2  becomes 
${1\over 2}\langle M\rangle^2_{a,b}(1+\sin^2\theta)+\langle M\rangle^2_c\cos^2\theta$ 
where $\langle M\rangle^2_{a,b}$  and $\langle M\rangle^2_c$ are the thermodynamic averages of the magnetization within the $a$-$b$ plane and along $c$-axis respectively. For the VF case the SF scattering is given by 
${1\over 2}\langle M\rangle ^2_{a,b} \sin^2\theta +\langle M\rangle^2_c \cos^2\theta$ 
and the NSF scattering is given by  $1/2\langle M\rangle^2_{a,b}$. The result is the same for the other reflections in the scattering plane if the appropriate value of $\theta$ is used. 	

In order to confirm the nature of the magnetic scattering in the antiferromagnetic parent compound and to check the operation of the spectrometer, measurements were made on IN20 with a large crystal of YBa$_2$Cu$_3$O$_{6.15}$. This material is an antiferromagnetic insulator with a transition temperature well above room temperature and has been studied previously \cite{tranquada,shamoto}.  Fig.\ 3a and b give the HF scattering for the first antiferromagnetic Bragg peak,  and show that the  signal all occurs in the SF channel and is thus totally magnetic. 
For the VF case shown in Fig.\ 3c and d the magnetic scattering is divided up between the SF and NSF channels. The sum of these should be the HF SF result. The least squares gaussian fits give $239\pm 5$ for HF SF scattering, 
$39\pm 1$ for the VF SF scattering and $204\pm 3$ for the VF NSF case. The sum is thus correct within the experimental error. The moment was 
earlier determined to lie in the $a$-$b$ plane, but the VF data can be used to check that result. 
$\theta=24.94$ degrees for the $(1/2, 1/2, 1)$ reflection so that the HF SF scattering would become 
$0.5889\langle M\rangle^2_{a,b}$ assuming $\langle M\rangle^2_c =0$. 
The VF SF scattering would be $0.0889\langle M\rangle ^2_{a,b}$ or 6.62 times smaller than the HF SF result. 
The measurement gives $6.13\pm 0.1$ which is slightly smaller than expected. The VF NSF scattering would be 
$0.5\langle M\rangle ^2_{a,b}$ so that the expected HF SF to VF NSF ratio would be 1.18 with the 
measured result being $1.17\pm 0.03$.  The data are consistent with the moment being essentially in 
the $a$-$b$ plane. Other reflections also can be used to confirm this result, but the 
$(1/2, 1/2, 1)$ reflection is the best, as the small size of the VF SF scattering makes a sensitive determination possible. 

\begin{figure}
\includegraphics[keepaspectratio=true, width=0.8\columnwidth,clip]{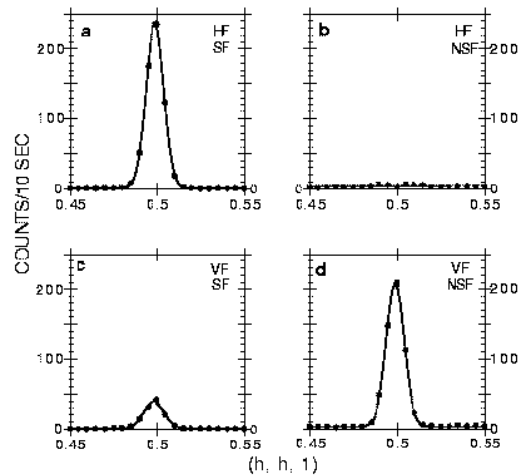}
\caption{
Polarized beam scattering from the antiferromagnetic insulator YBa$_2$Cu$_3$O$_{6.15}$. 
Both SF and NSF data are shown for the $(1/2, 1/2, 1)$ reflection. The solid lines are 
least squares fits to the data.
}
\end{figure}

\section{Sample Characterization}
The high quality single crystal of YBa$_2$Cu$_3$O$_{6.6}$ has been used in a number of 
neutron scattering experiments. This crystal provides excellent neutron scattering patterns. 
Presumably one of the reasons for this is that the chain order is well developed. The superconducting transition is narrow, being about 1.5K (10\%-90\%) as shown in Fig.\ 4a \cite{dai96}. The neutron scattering data obtained to characterize the chain order in Fig.\ 4b were acquired at the HB-3 spectrometer at the high-flux-isotope reactor 
(HFIR) at Oak Ridge National Laboratory. The data in Fig.\ 4c and d were obtained at the National Institute of Science and Technology (NIST) reactor using the SPINS spectrometer. The scan in Fig.\ 4b shows that the only 
chain order is of the ortho-II type and that the peak from this order at 0.5 is very intense. 
Following Andersen {\it et al.} \cite{andersen}, we fit the scattering from the oxygen superlattice by
$$
S(q)={A\over [1+(q_h/\Gamma_h)^2+(q_l/\Gamma_l)^2]^y},       
$$
where $\xi=1/\Gamma$ is the correlation length and $q$ is the deviation from the ordering modulation vector. 
Since the crystal is twinned we lump the  $a$ and $b$ directions together under the index $h$ while the 
$c$ correlatons are indexed by $l$. The least-squares fit in Fig.\ 5c yields a correlation length of 
$35\pm 1$ \AA\ along $a,b$ with $y=1.68\pm 0.08$ as expected for $3D$ domains with sharp boundaries. 
The fit of all the peaks along $c^\ast$ in Fig.\ 4d gives a correlation length of $23\pm 1$ \AA\ along $c$. 
This is very well defined chain order despite the fact that the crystal is twinned. The background under the $c$-axis reflections is also found to be very flat. A sloping background that is largest at $(0.5, 0, 0)$ is commonly found for less well ordered samples.  

\begin{figure}
\includegraphics[keepaspectratio=true, width=0.8\columnwidth,clip]{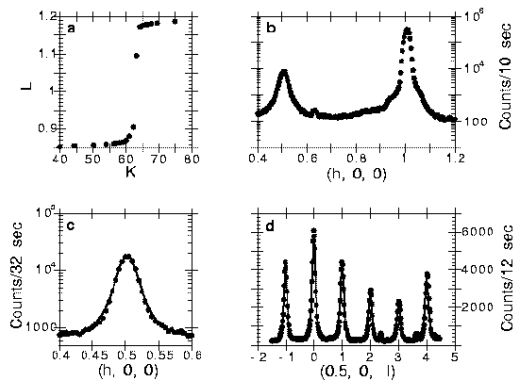}
\caption{
Characteristics of the YBa$_2$Cu$_3$O$_{6.6}$ crystal. a) shows the superconducting transition as measured by the inductance of the whole crystal.  A scan along the ${\bf a^\ast},{\bf b^\ast}$ 
direction made on the HB-3 spectrometer at HFIR  is shown in b). The peak for the ortho-II chain order 
is clearly seen at 0.5 while no peaks from any other type of chain order are observed.  A fit to the 
ortho-II peak along ${\bf a^\ast},{\bf b^\ast}$ is shown for data obtained on SPINS at NIST in c), while d) shows the fit for the peaks along ${\bf c^\ast}$ obtained at SPINS.
}
\end{figure}

This sample was the one used to discover incommensurate scattering in the YBa$_2$Cu$_3$O$_{6+x}$ system \cite{dai98,mook98} and to study the absolute intensity and temperature dependence of the resonance \cite{dai99}. The resonance peak is considerably narrower than that measured on other samples \cite{dai01} . The fact that only ortho-II chain ordering is observed might suggest that the oxygen concentration of our sample is somewhat lower than 6.6. This is possible although the temperature treatment used is expected to result in an oxygen concentration of 6.6, and the lattice constants of 
$a=3.826$ \AA, $b=3.876$ \AA, and $c =11.73$ \AA\ are those expected for YBa$_2$Cu$_3$O$_{6.6}$. 
This means that some of the normally empty chains of the ortho-II structure are at least partially filled. The sample contains Y$_2$BaCuO$_5$ in powdered form in small pockets in the crystal. The Y$_2$BaCuO$_5$ pockets are necessary to provide pathways for oxygen so that a very uniform oxygen concentration can be achieved.  However, these powdered phases produce rings in the scattering which generally make accurate measurements of the magnetic reflections difficult without polarization analysis. The sample weighs 25.6 grams and gives a counting rate of about $5000$ counts per second for the rather weak $(0, 0, 2)$ reflection on the IN20 spectrometer under the conditions used for the polarized beam experiment.

\begin{figure}
\includegraphics[keepaspectratio=true, width=0.8\columnwidth,clip]{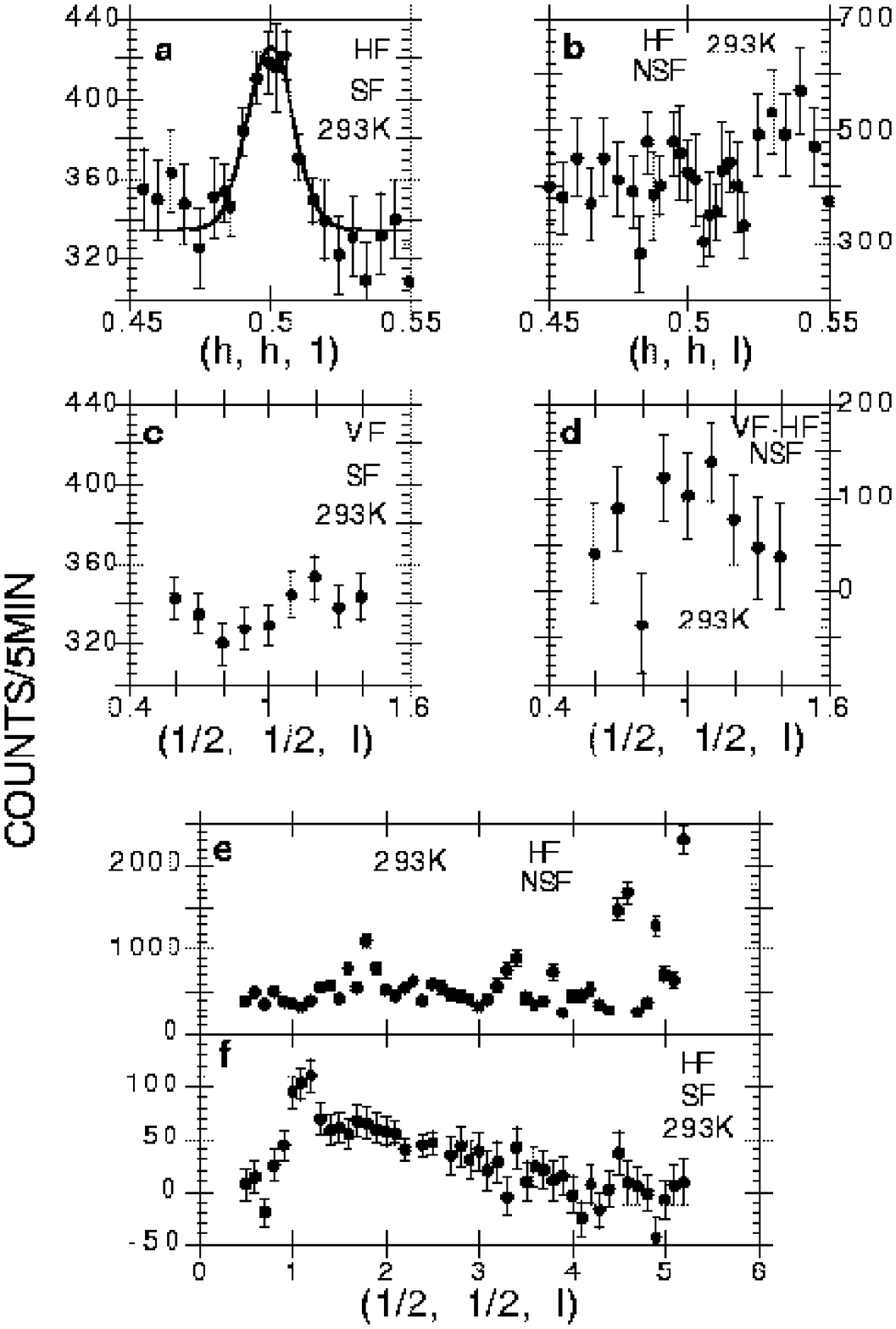}
\caption{
Polarized beam measurements made at 293 K. a) shows a HF SF scan through the $(1/2, 1/2, 1)$ reflection. 
The line is a Gaussian fit to the data. The HF NSF scan in b) shows no peak so that the peak in a) is completely magnetic. c) gives the result of a VF SF scan through the $(1/2, 1/2, 1)$ peak made along ${\bf c^\ast}$. d) 
is the same scan made in the NSF channel. In this case the data are the difference between the VF and HF field configuration. e) and f) are the NSF HF and SF HF results of a scan along the ${\bf c^\ast}$ direction.
}
\end{figure}

\section{Scattering in Region $C$}
The temperature dependence of the magnetic scattering in Fig.\ 1 has three main regions. 
The first region to be investigated is the signal that remains at the highest temperatures.  
The temperature dependence shown is from the $(1/2, 1/2, 2)$ reflection, however, the dependence 
observed with the $(1/2, 1/2, 1)$ reflection appears to be identical. The plot shows the signal minus 
background obtained $0.05$ rlu from both sides of the peak. Measurements were made at 293 K in region $C$ so as to be well removed from region $B$. The results of the measurements are shown in Fig.\ 5. 
From this point all the neutron spectra will be shown normalized to a 5 minute counting time to make it 
easy to compare the results from figure to figure. The error bars result from the total counting time used. Measurements made on the $(1/2, 1/2, 0)$ reflection showed no peak. This means that only antiferromagnetic coupling is present. The HF SF and HF NSF scans through the $(1/2, 1/2, 1)$ Bragg peak are shown in Fig.\ 5a and b and are similar to those of the YBa$_2$Cu$_3$O$_{6.15}$ insulator except very much smaller. The Gaussian fit gives a height of $91.3\pm 10$ 
counts and a width of $0.019\pm 0.003$ rlu HWHM.   The NSF scan shows some extraneous structure from nuclear scattering processes, particularly near 0.54 rlu. However, there is no peak at $(1/2, 1/2, 1)$ so the scattering there is purely magnetic. A HF SF and NSF scan along $c^\ast$ is shown in Fig.\ 5e and f. The NSF scan again has peaks that stem from extraneous nuclear scattering events. The SF scan showing the magnetic scattering has only one clear peak. This peak appears to be at or near to the $(1/2, 1/2, 1)$ position. A VF SF scan through this peak is given in Fig.\ 5c.  Any peak in this scan is certainly smaller than 20 counts. If the moment direction is in the $a$-$b$ plane we would only expect a peak about of $91.1/6.1$ or about 15 counts. This is obtained using the measured ratio for YBa$_2$Cu$_3$O$_{6.15}$. In this case the intensity of about 91 counts should show up in the VF NSF channel. Because the NSF scattering contains extraneous nuclear scattering and the magnetic signal is rather small the ${\rm VF-HF}$ difference is shown. This is on the order of the 90 counts as expected. It is clear the moment in region $C$ is located essentially in the $a,b$ plane just as is the case for the YBa$_2$Cu$_3$O$_{6.15}$ insulator. The peak width along $(1, 1, 0)$ is also the same value as for the insulator which is resolution limited. Defining a correlation length as $2\pi/FWHM$ would result in a correlation length of 
about 200 \AA. The width of the peak  along $c^\ast$ in Fig.\ 5f from a gausssian fit over the range 
from 0 to 1.4 rlu is $0.45\pm 0.05$ rlu. This would result in a correlation length of about 26 \AA. 

A pattern of magnetic scattering such as found in Fig.\ 5e is that expected from a magnet that is positionally disordered off the lattice sites of the crystal. This moment appears to derive from an impurity. Although we do not know definitely the nature of the impurity, it most likely is hydrogen. Analysis by high-resolution mass spectrometry of 
pieces removed from our crystals shows the number of any impurities is small, except for hydrogen. We note that the SF background for both HF and VF is about 340 counts. Since the background is equal in both cases, 
it can not come from random magnetic impurities and must stem from the nuclear spins in the crystal \cite{moon}. 
The NSF scattering is also in the neighborhood of 340 counts so the ratio of the SF to NSF background scattering is about 1:1. It is found for other crystals of YBa$_2$Cu$_3$O$_{6+x}$ that the ratio of the SF to NSF background scattering is about 1:4. Obviously the crystal contains a considerable amount of an impurity that has a large nuclear SF cross section. The material with the largest nuclear SF cross section is hydrogen, which has a nuclear incoherent SF scattering of almost 80 barns. 
It appears that the sample contains hydrogen as an impurity. 
It is likely that hydrogen enters the vacant chains of the ortho-II structure and in turn causes magnetism to appear in the planes.

Recent work shows that impurity derived antiferromagnetism from Co in the chains takes place even in the 
highest $T_c$ cuprate superconductors \cite{hodges} .  We do not know the exact structure of the impurity phase, but information on hydrogen based antiferromagnetism is given in Ref. \cite{macfarlane,gunther} . We have attempted to investigate this process by introducing water vapor into crystals with the ortho-II chain structure. The crystals took up a considerable amount of water, but showed no sign of any additional magnetic scattering at 
$(1/2, 1/2, 1)$. It is thus not clear exactly how the impurity driven antiferromagnetism takes place. The antiferromagnetism does appear to be closely related to that in the insulating phase with a high transition temperature and an $a,b$ plane ordered moment \cite{tranquada,shamoto}. The phase has long range correlation's along $a,b$ but is disordered with the lattice along $c$. It may have a slightly different $c$-axis lattice constant than the YBa$_2$Cu$_3$O$_{6.6}$ sample. Region $C$ is defined by the expected temperature dependence for the Ne${\rm \acute{e}}$l antiferromagnet, $A(1-T/T_N)^{0.5}$ where $A$ is a constant determined so that the curve passes through the 293 K data points. Since the Ne${\rm \acute{e}}$l temperature $T_N$ of the impurity phase is not known it was chosen to be 350 K. The shape of the curve is not greatly sensitive to $T_N$ and we did not want to 
risk destroying the crystal chain order by heating beyond 293 K. It seems likely the scattering observed by 
Sidis {\it et al.} \cite{sidis} is similar to this impurity phase. In fact the scattering in this case is much larger then that observed here. They suggest a moment of about 0.05 $\mu_B$ compared to the 0.02 $\mu_B$ 
determined earlier for the YBa$_2$Cu$_3$O$_{6.6}$ sample, which results in about 6 times more scattering. 
We have not observed this magnitude of magnetic moment in any crystal with a hole doping larger than 
YBa$_2$Cu$_3$O$_{6.35}$. It may be that the YBa$_2$Cu$_3$O$_{6.5}$ composition, which has empty chain spaces  in the ortho-II structure, is very susceptible to the introduction of impurities. 

\begin{figure}
\includegraphics[keepaspectratio=true, width=0.8\columnwidth,clip]{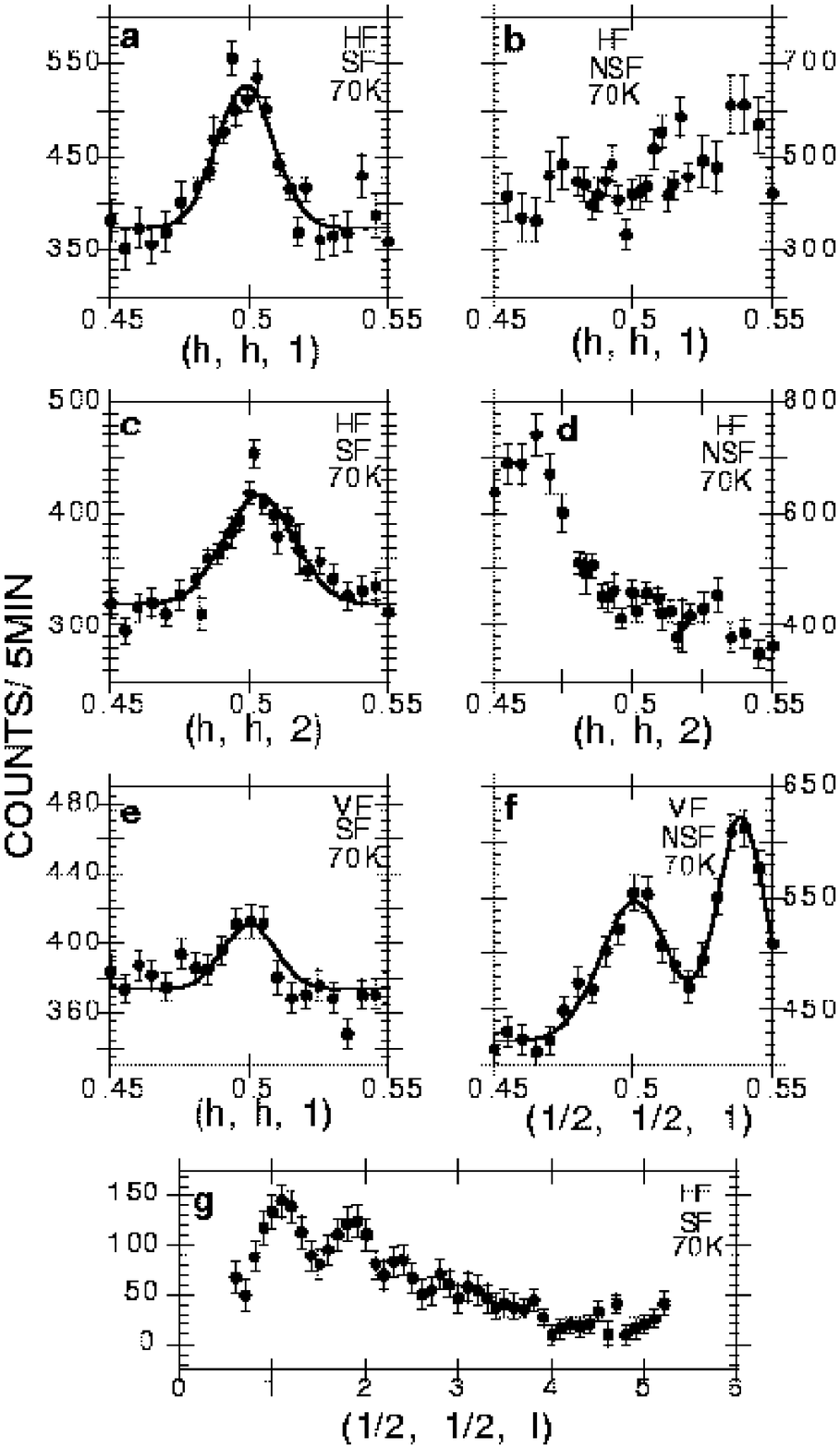}
\caption{
Polarized measurements made at 70 K. a) and b) show the HF SF and NSF scattering at the 
$(1/2, 1/2, 1)$ reflection just as is shown for the 293 K scattering in a) and b) of Fig. 5.  c) and d) 
give the same information for the $(1/2, 1/2, 2)$ reflection. Again the NSF scattering shows the 
influence of extranious nuclear scattering. e) gives the VF SF result for the $(1/2, 1/2, 1)$ 
reflection and f) is the VF NSF result. The line in f) is the least squares fit to two Gaussians. g) 
shows the result of a HF SF scan along the ${\bf c^\ast}$ direction.
}
\end{figure}

\section{SCATTERING IN REGION $B$}

The main increase in scattering defining region $B$ occurs somewhat below 200 K or at 
about $T^\ast$ for the YBa$_2$Cu$_3$O$_{6.6}$ composition. This is the region of current interest 
as the DDW state should become evident below $T^\ast$. The curve defining the region 
is similar to that established earlier \cite {mook1} except that the polarized measurements permit an accurate determination of the background which is difficult in the unpolarized case. Fig.\ 6 shows measurements made 
at 70 K which is the region of highest intensity, but still above $T_c$. The magnetic $(1/2, 1/2, 1)$ and 
$(1/2, 1/2, 2)$ reflections shown in Fig.\ 6a and b are wider than the $(1/2, 1/2, 1)$ reflection in region 
$C$, being $0.024\pm 0.001$ rlu  and $0.028\pm 0.001$ rlu HWFM respectively. The peak is likely the sum of a wide distribution from region $B$ and a narrow one from region $C$, however, the signal is too small to try to distinguish these two distributions. Nevertheless, the correlation length is clearly shorter in the $B$ phase. The correlation length in the $c^\ast$ direction is also shorter in region $B$ based on the width of the 
$(1/2, 1/2, 1)$ reflection shown in Fig.\ 6g. A Gaussian fit over the range from 0 to 1.4 rlu gives 
a width of $0.82\pm 0.15$ rlu HWHM for the 70 K data. This results in correlation lengths of 
about 150 \AA\ from the $(1/2, 1/2, 1)$ peak and 14 \AA\  along $c^\ast$. Again no $(1/2, 1/2, 0)$ 
peak is observed so that the magnetic structure is antiferromagnetic.

The shorter correlation length plus the increase in intensity identifies a new phase since a 
broadening by itself would decrease the intensity. Fig.\ 6b and d show no peak in the NSF intensity at the Bragg position so that the observed peak is entirely magnetic in nature. Fig.\ 6e gives the VF SF scattering and provides 
one of the most important results. If the moment direction is the same as for the antiferromagnetic insulator  
this peak should  be 6.1 times smaller than the scattering in Fig.\ 6a and 6g or $25\pm 1$ counts. 
Instead the fit in Fig.\ 2e gives $38\pm 5$ counts. The VF NSF scattering is shown in Fig.\ 6f. This is contaminated by extraneous nuclear scattering near 0.54 rlu as can also be seen from the HF data in Fig.\ 6b. 
However, the region near 0.5 rlu is reasonably clear of extraneous scattering so two Gaussian distributions have been fitted to the data. The height of the distribution at 0.5 rlu is $123\pm 7$ counts. The ratio of VF NSF to HF SF scattering is thus $4.9\pm 0.3$ where it should be should be 5.6 for $a,b$-axis moments. The more accurate comparison is between the HF and VF SF data, especially as it is free from extraneous nuclear scattering.  A $c$-axis component of the moment direction is thus found for the scattering in region $B$. In fact an appreciable part of the increase in intensity attributed to region $B$ results from $c$-axis directed moments.  This can mean that the scattering in region $C$ acquires a $c$-axis moment component at lower temperatures. However, it is more likely that an appreciable part of the additional moment in region $B$ is directed along the $c$-axis. Fig.\ 6g shows that the increase in intensity comes at positions associated with the YBa$_2$Cu$_3$O$_{6.6}$ lattice as can be particularly noted for the $(1/2, 1/2, 2)$ reflection. The peaks in the curve are not exactly at the reciprocal lattice points, but the new intensity which is added to that shown in Fig.\ 3f does occur on the lattice points.

The scattering in region $B$ is thus quite different than that in region $C$ thought to result from an impurity phase. The peaks in the scattering in the $B$ phase stem from Bragg scattering from the YBa$_2$Cu$_3$O$_{6.6}$ lattice. Of course the scattering in $B$ could come from a different type of impurity phase, but this would have to have a moment direction that is appreciably directed along the $c$-axis. No such impurity scattering has been previously observed.  The correlation lengths of the moments in the $B$ phase are also considerably shorter that in the $C$ phase. The work by Sidis {\it et al.} \cite{sidis}  does not show a distinct $B$ phase. However, the scattering from the impurity phase is so strong in their case that the $B$ phase would go unnoticed. 

The $A$ phase where increased scattering is observed below $T_c$ has not been thoroughly investigated. 
The ratio of HF SF scattering to VF SF scattering does not seem to be appreciably different than for region 
$B$ and the peak widths do not change noticeably. Since such a result has also been seen 
by Sidis {\it et al.} \cite{sidis} it would seem that the measurements are observing the same effect. However, the origin of the increased scattering is unknown.

\section{THE HIGHER ANGLE REFLECTIONS}
In general the pattern of the reflections can be used to determine the moment direction. Also the size of the outer reflections should help determine whether the moment stems from electron spins or from 
orbital moments such as the DDW state. However, the situation is complicated by the presence of more 
than one kind of order. A reasonable picture of the peaks in the $B$ phase can be obtained by subtracting the data taken at 293 K  from the 70 K data. Fig.\ 7a shows this result smoothed to obtain a result that is easier to visualize. Smoothing can introduce artificial structure if the data contains a spurious data point, but the results in Fig.\ 5f and 6g appear well behaved. In general SF data are free from spurious effects. A case to the contrary is shown in c, d, and e of Fig.\ 7, but this can be easily identified as the NSF data contains a huge peak that likely stems from higher order scattering. The main result desired from Fig.\ 7a is the ratio of the heights of the Bragg peaks.  Essentially the same result is obtained by subtracting the data and adding the 5 points surrounding the various Bragg positions together. The line shown is a Gaussian fit to the peaks at the 5 positions along $l$. The fit at the $l=1$ and 2, positions was unconstrained while at $l=3,$ 4, and 5 the fit was constrained to the lattice positions with a width set the same as for the $l=2$ case.

\begin{figure}
\includegraphics[keepaspectratio=true, width=0.8\columnwidth,clip]{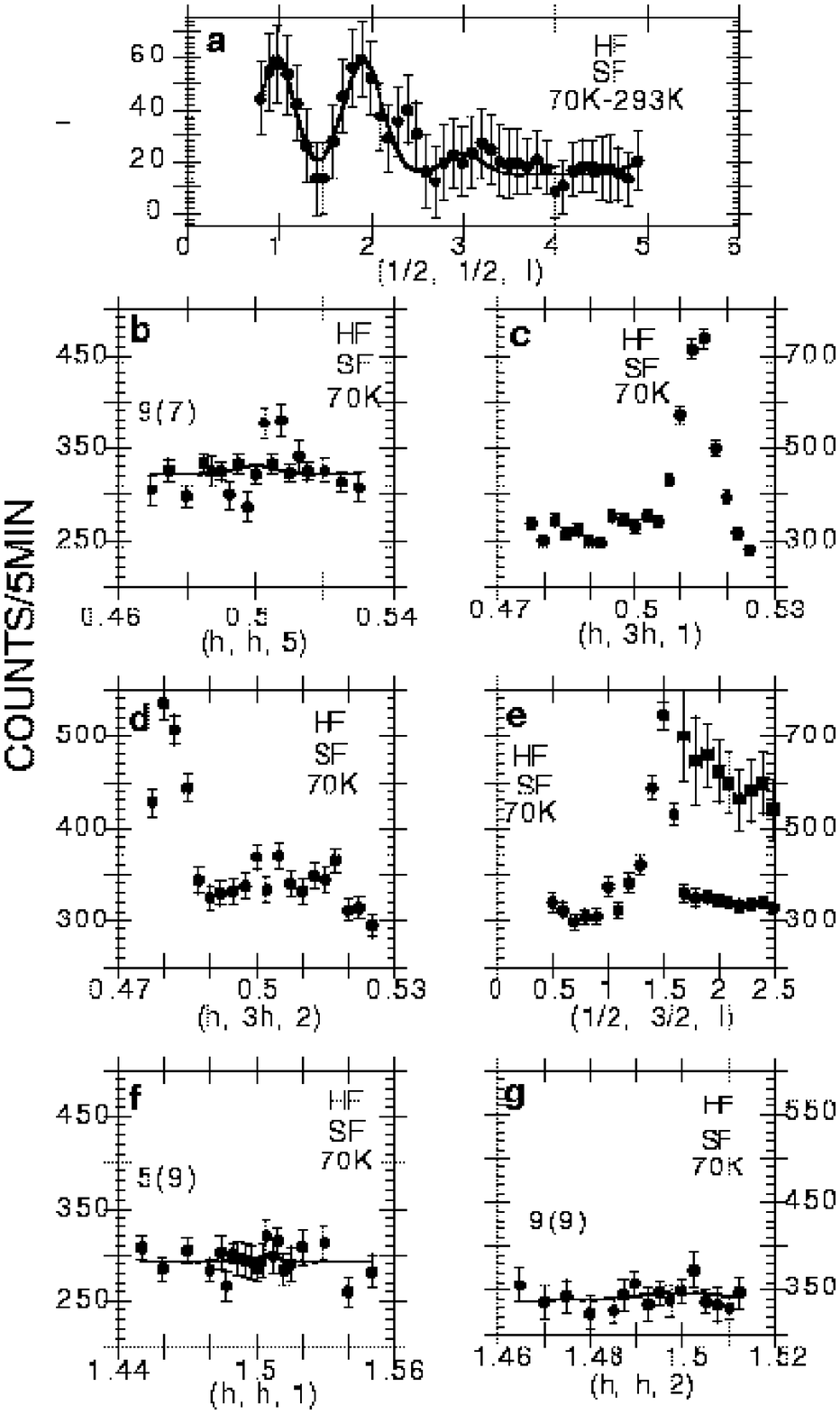}
\caption{
Comparison of the lower angle and higher angle HF SF reflections in the $B$ phase. 
 a) gives the 
result of subtracting the $l$ dependent data at $(1/2, 1/2, l)$ at 293 K from 70 K data. 
The result has been smoothed over 0.3 rlu.  b) is a $(h, h, 5)$ scan at 70 K. The line is a 
gaussian fit with the center fixed at $(1/2, 1/2, 1)$, and the width set at 0.014 rlu, which is the measured spectrometer resolution at this position. c) and d) are  $(h/2, 2h/3, 1)$ and $(h/2, 2h/3, 2)$ scans.  
These contain a large spurious peak that stems from a nuclear scattering event visible through second 
order scattering. e) gives the result of a $(1/2, 3/2, l)$ scan and again contains a large spurious nuclear peak. 
The square points on the right of the graph show the data multiplied by 5 with a constant subtracted so the points remain on the plot.  f) and g) are scans through the $(3/2, 3/2, 1)$ and $(3/2, 3/2, 2)$ peaks. The line shows a fit made in the same way as for b).
}
\end{figure}

The peak ratios depend on the magnetic structure, form factor, and spectrometer resolution. These can be calculated in a straightforward way for the antiferromagnetic insulator. However, we have measured them for the YBa$_2$Cu$_3$O$_{6.15}$ material and the ratios for the $l=1$ through 5 peaks normalized to $l=1$ are 1, 1.18, 0.13, 0.06, 0.82. These agree with the expected values for the previously determined magnetic structure \cite{tranquada}. The most obvious difference between these and the result in Fig.\ 7a is the lack of a peak at $l=5$. This can be seen as well from the unsubtracted data in Fig.\ 6g. The peak ratio determined from Fig.\ 7a is 1, 0.89, 0.17, 0, 0.10. The $l=5$ peak is also shown in Fig.\ 7b in a scan along $(h, h, 5)$. The fit gives a peak of $9\pm 7$ counts or a very small number. This scan is across the ridge of scattering that results from region $C$ so it includes any scattering that might stem from this phase. However, Fig.\ 5f shows any contribution from the $C$ region is small.   We also have an estimate for the size of the $l=1$ peak from the difference of the scans in Fig.\ 6a and Fig.\ 5a. This difference is about 61 counts which is larger than that in Fig.\ 7a, but not inconsistent with it since the peak intensity is reduced by the smoothing process. However, it confirms the small size of the $l=5$ peak compared to the $l=1$ peak. Zero on Fig.\ 7a is about 15 counts demonstrating the small background increase at 70 K.  

It is clear that the outer reflections are small or nonexistent for both the $C$ and $B$ phases. However, the $C$ phase shows only one peak and a mostly featureless ridge of scattering. This is the signature of a positionally disordered phase in which the outer reflections are spread out so much as to be invisible. The $B$ phase has two clear reflections of comparable width centered on the magnetic lattice. This is not the signature of a spatially disordered magnet so that the smallness of the outer reflections must stem from another source.  In fact the pattern is quite similar to that expected for $c$-axis order as characterized by an intensity ratio for the $(1/2, 1/2, l)$ reflections of 1, 0.90, 0.12, 0, 0.02. This serves to confirm the VF polarized result for region $B$. It is also similar to the pattern expected for orbital moment scattering \cite{chakravarty01a,chakravarty01b} . However, the data are not sensitive enough to distinguish between the two cases.

If the moment in region $B$ is a Cu spin moment, a noticeable intensity should be visible in the next magnetic zone out along  $[1, 3, 0]$ which is the $(1/2, 3/2, l)$ zone. The intensity of the $(1/2, 3/2, 1)$ and $(1/2, 3/2, 2)$ reflections should be 0.44 and 0.5 of the $(1/2, 1/2, 1)$ reflection respectively assuming a $3d_{x^2-y^2}$ 
Cu spin form factor. These reflections are shown in Fig.\ 7c-e. Unfortunately, the large higher order nuclear peak that contaminates the data makes the determination of a small peak difficult. The expected peak of about 31 counts is not obvious in c and d, but it is difficult to rule it out. There also may be some contribution from region $C$ in these scans. The scan along $l$ in e is more useful as a $(1/2, 3/2, 2)$ peak would not be expected from region $C$ since there is no $(1/2, 1/2, 2)$ peak. Since the inset with the square points has been scaled up by a factor of 5, 
a peak of about 150 counts is expected. There is no indication of a peak of this size. This is in favor of an orbital moment interpretation of the moment in region $B$ since the form factor would be expected to fall off more quickly in the orbital moment case. Data obtained in the $(3/2, 3/2, l)$ zone are included in f and g for completeness. The fit is again a gaussian fixed at the magnetic Bragg point with the width fixed at the expected value determined from the spectrometer resolution. The peak heights determined are $5\pm 9$ and $9\pm 9$ for the $(3/2, 3/2, 1)$ and $(3/2, 3/2, 2)$ reflections. These are small, however, the expected peak heights from Cu spin magnetism for the moment attributed to the $B$ phase are about 10 counts or comparable to the experimental errors.

\section{INELASTIC SCATTERING AND THE GAP}
Measurements of inelastic scattering to determine the excitations in the gap have also been made. Results on a low energy scale were obtained at the SPINS spectrometer at NIST while results at higher energies were obtained on the HB-3 triple-axis spectrometer at HFIR. The data are presented in Fig.\ 8. The scan in Fig.\ 8a is unpolarized so that structure from nuclear contamination is found. Nevertheless a peak over background of about 2000 counts is observed The scan is made with a neutron energy of 5 meV and open beam collimation. Two Be filters in the beam removed higher order contamination. Fig.\ 8b shows an energy scan demonstrating that no inelastic signal over the 20 count background is found.  Fig.\ 8c is a scan for a 1 meV transfer. Any excess signal at the magnetic position is smaller than about 2 counts. The gap is extraordinarily clean. The elastic signal in Fig.\ 8a is estimated to correspond to about 0.02 
$\mu_B$ and any signal at 1 meV is $10^{-4}$ times smaller.  No excitations stemming from the elastic signal are thus observed. This is not unexpected as spin excitations from the $C$ phase are likely to have a 
gap larger than 1 meV. Any orbital scattering from the DDW state should have a gap comparable to the pseudo gap. 
Fig.\ 8d shows measurements made on the HB-3 spectrometer. Pyrolytic graphite was used for the monochromator and analyzer and the collimatiom was $40^\prime$-$40^\prime$-$80^\prime$-120$^\prime$ from in front of the monochromator to behind the analyzer. The final energy was held fixed at 30.5 meV and a pyrolytic graphite filter was used in the scattered beam. The measurements were made around the $(1/2, 3/2, 1.8)$ magnetic position.  No peak is observed at the magnetic position at 10 K and 70 K. Magnetic scattering begins to be observed at 150 K as the pseudogap temperature is reached. Note from Fig.\ 1 that this is the temperature where most of the scattering in region B is lost. The data show the extreme cleanness of the gap at 10 K and the pseudogap at 70 K. Fig.\ 8e shows the 34 meV resonance as determined from temperature subtracted data near the same magnetic position used in Fig.\ 8d. The height of the resonance is about 400 counts over background. The scattering at 10 meV and 70 K can be at most 5 counts. The fit gives $-13\pm 9$ counts at 70 K. The scattering in the pseudogap at 10 meV is thus at least 50 times smaller than the resonance. This very clean gap is another signature of the quality of the sample. 

\begin{figure}
\includegraphics[keepaspectratio=true, width=0.8\columnwidth,clip]{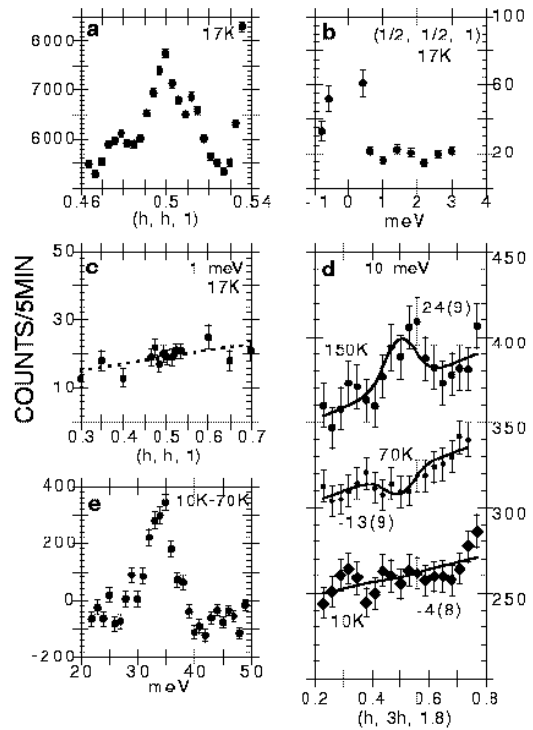}
\caption{
Inelastic measurements to examine the gap. a) shows an unpolarized scan of the $(1/2, 1/2, 1)$ 
reflection on SPINS.  b)  and c)  show inelastic scans demonstrating that no excitations are visible. d) gives a measurement on HB-3 at the magnetic position for energy transfer of 10 meV. The lines are Gaussian fits to the data. The fit at 150 K is an unconstrained fit of a gaussian on a sloping background. Fits at 70 K and 10 
K were constrained at 0.5 with a width identical to that obtained at 150 K. The results of the peak heights are given with the errors in parenthesis. 50 counts were added onto the 70 K data and 75 counts to the 150 K data so that the results could be compared on the same plot. e) gives the resonance at the same magnetic position used for d).
}
\end{figure}

\section{magnetic field dependence}
The magnetic field dependence of the elastic scattering at the $(1/2, 1/2, 1)$ magnetic Bragg point has been made on SPINS. In this case the crystal was orientated so the field was approximately along the $c$-axis \cite{dain}. 
This means the beam must pass through the thick part of the sample crystal so that the overall signal is smaller than for the measurements in Fig.\ 8a. The results are shown in Fig.\ 9. A scan at 30 K with and without the 7 T 
applied field is shown in Fig.\ 9a. It is obvious that there is no sizable field effect. 
The temperature dependence of the scattering at 0 T and 7 T is shown in Fig.\ 9b. Any field dependence is small, but if there is any effect the scattering appears to get larger in the field. That might suggest the signal increases as vortices are introduced into the crystal. However, the increase seems to appear above $T_c$ as well as below and a small shift in the crystal as the field is applied may account for the effect. The scattering in region $C$ attributed to a high temperature spin ordering would not be expected to change appreciably in a 7 T field. Likewise scattering from orbital ordering would be expected to insensitive to fields of this size \cite{nguyen,marston}. It would be interesting to make measurements in larger fields and these will be attempted at a later date.  

\begin{figure}
\includegraphics[keepaspectratio=true, width=0.8\columnwidth,clip]{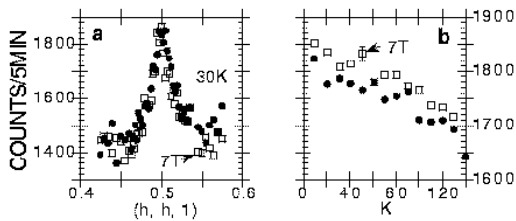}
\caption{
Measurements made on the SPINS spectrometer on the magnetic field dependence of the elastic magnetic intensity. The field was applied along the $c$-axis of the crystal. a) shows a scan made through the magnetic peak at 0 and 7 T. 
b) gives the temperature dependence of the scattering at the magnetic peak position with and without 
the 7 T field. In a) and b) the data with the field on are the open squares.
}
\end{figure}

\section{other sample crystals}
\begin{figure}
\includegraphics[keepaspectratio=true, width=0.8\columnwidth,clip]{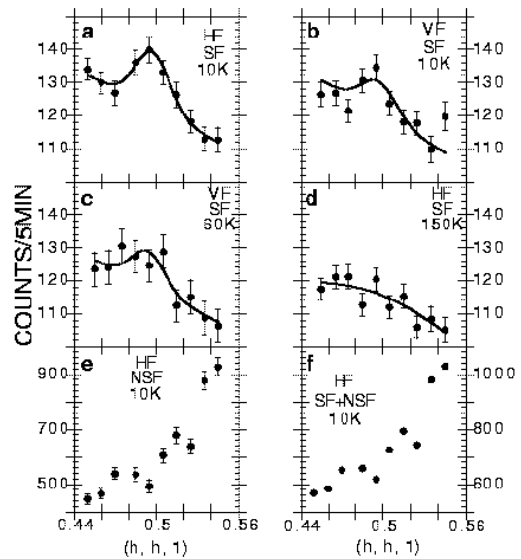}
\caption{
Polarized neutron data for a crystal of YBa$_2$Cu$_3$O$_{6.45}$.  a), b), c), and d) 
show HF and VF results for different temperatures for the $(1/2, 1/2, 1)$ reflection. 
The solid line is a gaussian fit to the data. The fit is unconstrained for a),  but is constrained to occur at (1/2, 1/2, 1) for b), c), and d). e) gives the HF NSF scattering at 10 K while f) is the sum of the scattering in a) and e).
}
\end{figure}

Polarized beam measurements were made on a number of other sample crystals of different compositions. In each case the signal from magnetic order was either absent within the errors of the measurement or much smaller than the scattering from the YBa$_2$Cu$_3$O$_{6.6}$ crystal. The crystals were all about the same size except for a 
YBa$_2$Cu$_3$O$_7$  crystal which weighed about 80 grams. This crystal showed no observable magnetic signal. 
This is not unexpected for DDW scattering which would have a small intensity at high hole doping levels. A small signal was observed for crystals with the composition YBa$_2$Cu$_3$O$_{6.7}$ and YBa$_2$Cu$_3$O$_{6.45}$. 
$T_c$ for these crystals is 74 K and 48 K respectively.  These crystals are the same ones characterized 
in Ref. \cite{dai01}, and scans showing the incommensurate and resonance scattering are available there. 
The crystals are not as highly ordered as the YBa$_2$Cu$_3$O$_{6.6}$ crystal with the superconducting 
transition being 2.7 K wide for the YBa$_2$Cu$_3$O$_{6.7}$ sample and 5 K wide for the YBa$_2$Cu$_3$O$_{6.45}$ 
crystal. The resonance and incommensurate structure are much less well defined than for the YBa$_2$Cu$_3$O$_{6.6}$ crystal. However, the ratio of the SF to NSF background scattering was about 1 to 4 showing a much smaller SF background. It appears likely the highly ordered ortho-II phase is very susceptible  to impurities. 

Data for the YBa$_2$Cu$_3$O$_{6.45}$ crystal are shown in Fig.\ 10. The HF SF scattering in Fig.\ 10a is 
clearly visible but is very small. The Gaussian fit yields a height of $17\pm 2$ counts and a width of 
$0.03\pm 0.005$ rlu.  The width is thus wider than the resolution and is of a similar width as the scattering 
in region $B$ for the YBa$_2$Cu$_3$O$_{6.6}$ crystal. The intensity is about 3 times 
smaller than the moment attributed to region $B$. The size of the VF scattering appears to be 
comparable with the HF scattering indicating a $c$-axis directed moment. The scattering diminishes 
as the temperature is increased. The NSF scattering is contaminated with spurious nuclear scattering as 
in the YBa$_2$Cu$_3$O$_{6.6}$ case. Fig.\ 10f gives the result of adding the strongest magnetic signal shown in 
Fig.\ 10a to the NSF scattering. It is obvious that no magnetic signal is observable without a polarized beam. The observed scattering is what might be expected from the DDW phase. However, the signal is so weak that it is hard to characterize it in any detail. We are now searching for a crystal that shows a stronger magnetic signal, but is free of the high SF background scattering found in the YBa$_2$Cu$_3$O$_{6.6}$ crystal. 

\section{Summary and Conclusions}
It is now obvious that searches for a small magnetic moment of the size expected for the DDW state are only accomplished with considerable difficulty. 
Small static moments are more difficult to observe than dynamics because of the large background from competing nuclear and magnetic scattering. Polarized neutrons appear to be a necessity to get a clear picture of the magnetic scattering. The problem of obtaining a sample crystal with sufficiently low disorder to observe the DDW state is perhaps the most difficult one. Samples with the ortho-II structure probably have the lowest disorder, but appear to be very sensitive to the introduction of impurity phases. The three experiments done on crystals of this type all show some type of extraneous magnetic signal. The Sidis {\it et al.} \cite{sidis} experiment shows a large moment that orders at high temperatures that must stem from an impurity since moments of this size are observed in no other samples of roughly the same composition. The Stock {\it et al.} \cite{stock} experiment shows a very large roughly $Q$ independent magnetic background that increases as the temperature is lowered that is not observed in other samples. Note that the background in Fig.\ 10a and b is essentially unchanged with temperature despite the very high sensitivity of the experiment. Our crystal of YBa$_2$Cu$_3$O$_{6.6}$ appears to have a high temperature impurity phase that is smaller, but similar to the Sidis {\it et al.} \cite{sidis} sample. It is interesting that in both cases the magnetic scattering 
increases below $T_c$.  This effect may be important in its own right and should be explored further. 
Any crystal to be used in such experiments obviously has to be carefully characterized. Obvious requirements are a very clean gap and pseudogap. 

Despite the difficulties our experiments show evidence that a state with an orbital moment may well exist. 
Such a state would seem to be best characterized by a moment that is predominately directed along the $c$-axis 
and appears at roughly $T^\ast$. The $B$ phase in our YBa$_2$Cu$_3$O$_{6.6}$ crystal appears to have these properties. Orbital moments should also have a form factor that falls off faster than the spin form factor. 
This also appears to be the case for the scattering from the $B$ phase. However, this determination is made difficult by the existence of the high temperature ordered phase. Care also has to be taken that the moment in question is ordered on the crystal lattice as a disordered moment such as that ascribed to the $C$ phase also has small outer reflections. 

Additional measurements are needed to clearly determine the existence of orbital moments and the DDW phase. It appears that these are best done by obtaining a crystal of the ortho-II phase without magnetic impurities. If this is impossible less well ordered crystals like the YBa$_2$Cu$_3$O$_{6.45}$ sample that show no evidence of impurity scattering may have to be used despite the small magnetic signal. $\mu$SR measurements like those of 
Sonier {\it et al.} \cite{sonier} can observe the effect of very small moments. However, the moment is difficult to characterize sufficiently well in such experiments to determine its origin. Neutrons can identify if the moments lie on the crystal lattice, the moment direction, and the magnetic form factor. This is essential information needed to identify orbital scattering. 

Finally it appears that the cuprates display a number of competing phases so that for instance stripe order may appear in some samples, charge order in others, and orbital order in others yet. It is thus possible to have experiments that seem to disagree, but are only sampling the order present in a particular sample. This makes sample characterization particularly important in experiments to determine the true nature of the pseudogap. 

\section{Acknowledgements}
We are grateful to S. Chakravarty, R. Laughlin, D. Morr, D. Pines, and B. Marston for helpful conservations. 
M. Enderle provide valuable assistance with the IN20 spectrometer. We thank W.J.L. Buyers for supplying his results prior to publication. 
This work was supported by U.S. DOE under contract 
DE-AC05-00OR22725 with 
UT-Battelle, LLC.

\pagebreak

\end{document}